\documentstyle{article}

\def\be{\begin{eqnarray}}
\def\ee{\end{eqnarray}}
\def\nn{\nonumber}

\hoffset= - 1.0in         

\begin{document}

\hfill ITEP/TH-21/08

\bigskip

\centerline{\Large{
NSR Superstring Measures Revisited
}}

\bigskip

\centerline{A.Morozov}

\bigskip

\centerline{{\it ITEP, Moscow, Russia}}

\bigskip

\bigskip

\centerline{ABSTRACT}

\bigskip

{\footnotesize
We review the remarkable progress in evaluating the NSR
superstring measures, originated by E.D'Hoker and D.Phong.
These recent results are presented in the old-fashioned form,
which allows us to highlight the options that have been
overlooked in original considerations in late 1980's.
}

\bigskip

\bigskip

\tableofcontents

\section{Introduction}

After the role of holomorphicity in $2d$ conformal theories
was fully realized and exploited in \cite{BPZ} it was
natural to look for the holomorphic factorization in the
conformal-invariant first-quantized theories of critical
strings \cite{PO}. The problem here was that the relevant
quantities had to be meromorphic not only in $z$-variables,
which define positions of operators in operator-product
expansions, but also in the moduli of Riemann surfaces.
The relevant holomorphic anomalies in Polyakov's combination
of determinants,
which define string measures for bosonic, super- and
heterotic strings, were evaluated in \cite{BK}
and shown to vanish together with conformal anomaly of \cite{PO}.
This Belavin-Knizhnik theorem became a starting point for
construction of perturbative string and conformal field
theories, reviewed, for example, in \cite{VV}-\cite{MPrev}.
Without Belavin-Knizhnik theorem the Polyakov string
measures could be discussed in terms of either Shottky
parametrization \cite{Shopar} or Selberg traces \cite{Selt}.
With this theorem the adequate language became that of the Mumford
measure $d\mu$ on the moduli space of complex curves
(= Riemann surfaces) \cite{Mum,YuM}:
the measure for bosonic string was proved in \cite{BK}
to be $\frac{|d\mu|^2}{\det ({\rm Im}\ T)^{13}}$, while that
for the NSR superstring \cite{superst}
had to contain an extra factor of
$\Big({\det\ ({\rm Im}\ T)\Big)^{8}}$
with $d\mu$ presumably multiplied by some
modular form $\Xi_8$ of the weight $8$
(in the case of heterotic string \cite{heter} this NSR measure
is multiplied by a complex conjugate of $d\mu$ times one of the
two similarly different but actually coinciding weight-eight
modular form, denoted by $\xi_8$ and $\xi_4^2$ below
\cite{BK,BKMP,M1,GM}).
The first big success on this way was explicit construction
of $d\mu$ for the genera $2,3$ and $4$ in terms of
period matrices in \cite{BKMP,M1,GM} -- and this was the
starting point of the long road towards DHP construction
of NSR measures  in \cite{DHP1}-\cite{PG}.

From the very beginning there were two related but different
strategies.

The first approach was to begin with Polyakov's measure for
NSR string at given characteristic $e$,
expressed through determinants in \cite{PO} and
holomorphically factorized in \cite{BK},
integrate away the "supermoduli" and obtain the relevant
modification $d\mu[e]$ of the Mumford measure.
This road looked straightforward \cite{Mar}-\cite{Ort},
until it was
shown in \cite{M2}-\cite{LN} that naive integration over
supermoduli does not work and its proper version requires a
lot of work. This work was finally done by Eric D'Hoker and
Duong Phong (DHP) in a  series of impressive papers
\cite{DHP1}-\cite{DHP9}, but only 15 year later and only for
genus $2$ so far.

The second equally obvious approach was to make educated
guesses for NSR superstring measure, i.e. to find the
relevant weight-$8$  modular forms from their expected
properties, at least for the first low genera, like it was
done in \cite{BKMP,M1,GM} for $d\mu$ itself.
As explained in \cite{MP1}, the main obstacle on this way
was modular non-invariance of the Riemann identities
-- which are necessarily used for cancelation of tachionic
divergencies after GSO projection
(=sum over characteristics) \cite{GSO}.
After a series of attempts \cite{att} -- now known to be
partly misleading
\footnote{
Note, however, that a lot of results from that period
remain quite important and actually
relevant for discussion of correlators on the lines of \cite{DHP8}.}
--  this approach was temporarily abandoned.
Now, after the DHP triumph it is used again and already led
to explicit construction of NSR measures
at genera $3$ \cite{CPG1}, $4$ \cite{Gru,CPG2} and
-- somewhat less explicitly -- for all higher genera \cite{Gru}.
The problem for $g>4$ is that the Mumford measure $d\mu$
does not possess any nice representation in terms modular forms
(only a far more transcendental formulas of \cite{Krev,M2,MPrev}
are currently available), but the result of \cite{Gru}
supports the original suggestion of \cite{BK,M1} that the
ratio $\Xi_8[e] = d\mu[e]/d\mu$ is a modular form
(then it has modular weight $8$) and this $\Xi_8[e]$ is
proposed in \cite{Gru} in a simple and clear form.
The only remaining problem with these suggestions at $g\geq 3$
is related to $1,2,3,4$-functions,
and this makes the story of NSR measures not fully completed.
Still, we already know quite
a lot, and the time probably came to analyze and explain the
failures of the early attempts and understand what are the
answers to the questions, posed but unanswered in late 1980's.
This paper is an attempt of such analysis.

\section{Riemann surfaces and theta constants \cite{Mumb}-\cite{Ig}}

\subsection{Theta-functions, theta-constants and modular forms
on the Siegel semi-space}

\subsubsection{Theta functions}

Theta-functions are special functions, associated with
abelian varieties: $g$-dimensional tori, which are factors
of $C^g$ over relations $z_i \sim z_i + T_{ij}z_j$, where
{\it symmetric} period matrices $T_{ij}$ with positive definite
imaginary part $({\rm Im}\ T)$ are points in the
$g(g+1)/2$ Siegel semi-space, defined modulo integer symplectic
(also called {\it modular})
transformations $T \sim (AT+B)/(CT+D)$ from the group
$\ {\rm Sp}(g,Z)$.

Bosonic and super-string measures on the moduli space
of Riemann surfaces are defined in terms of theta-functions
with semi-integer characteristics, this is taken into account
in the following definition:
\be
\theta\left[\begin{array}{c}\vec\delta\\ \vec\varepsilon \end{array}
\right]\big(\vec z|T\big)
= \sum_{\vec n \in Z^g} \exp \left\{
i\pi \Big(\vec n + \frac{1}{2}\vec\delta\,\Big)T
\Big(\vec n+ \frac{1}{2}\vec\delta\,\Big)
+ 2\pi i \Big(\vec n + \frac{1}{2}\vec\delta\,\Big)
\Big(\vec z+\frac{1}{2}\vec\varepsilon\Big) \right\}
\ee
Sums are over all $g$ vectors $\vec n$ with integer coordinates,
each coordinate of characteristic vectors $\vec\delta$ and
$\vec\varepsilon$ can take values $0$ or $1$.
Characteristic is called even or odd if scalar product
$\vec\delta\vec\varepsilon\ $ is even or odd respectively and
associated theta-function is even or odd in $\vec z$.
The value of theta-function at $\vec z=0$ is called theta-constant,
it automatically vanishes for odd characteristic.
We often denote characteristics by
$e = \{\vec\delta,\vec\varepsilon\}$, in most cases these will be
even characteristics, when we refer to {\it some} odd characteristic
it is labeled by $*$.
There are $N_e=2^{g-1}(2^g+1)$ even and $N_*=2^{g-1}(2^g-1)$ odd
semi-integer characteristics:
$$
\begin{array}{|c|c|c|}
\hline
g & N_e & N_* \\
\hline
1 & 3 & 1 \\
2 & 10 & 6 \\
3 & 36 & 28 \\
4&136& 120\\
\ldots&&\\
\hline
\end{array}
$$
With a pair of characteristics (not obligatory even)
we associate a sign factor
\be
<e_1,e_2>\ = \exp \big\{i\pi (\vec\delta_1\vec\varepsilon_2-
\vec\varepsilon_1\vec\delta_2)\big\} =
(\vec\delta_1\vec\varepsilon_2-
\vec\varepsilon_1\vec\delta_2) {\rm mod}\, 2
=\ <e_2,e_1>
\ee
which takes values $\pm 1$.
In particular, $<e,e>\ = 1$.

\subsubsection{Modular forms}

Functions of $T$, transforming multiplicatively under
modular transformations,
$f(T) \rightarrow \Big(\det (CT+D)\Big)^{-k} f(T)$,
are called modular forms of weight $k$.
Theta-constants are not modular forms, they are not
simply multiplied by $\Big(\det (CT+D) \Big)^{-1/2}$,
but also acquire additional numerical factors proportional
to $e^{i\pi/4}$ and change characteristics.

The simplest modular forms can be made from the $8$-th powers
of $\theta$-constants, since modular transformations act on
them just by permuting their characteristics.
In particular, for any integer $k$
and $g$
\be
\xi_{4k} \equiv \sum_e^{N_e} \theta_e^{8k}
\ee
is a modular form of weight $4k$.
Important for NSR measures are
\be
\xi_4 = \sum_e^{N_e} \theta_e^{8}\ \ \ \ {\rm and}\ \ \ \
\xi_8 = \sum_e^{N_e} \theta_e^{16}
\ee
Also
\be
\Pi \equiv \prod_e^{N_e} \theta_e
\label{Pif}
\ee
of weight $N_e/2$ is a modular form for $g\geq 3$,
while roots of unity arise and $\Pi$ should be raised
to power $8$ and $2$ at $g=1$ and $g=2$ respectively.
This $\Pi$ is the building block of Mumford measure
at $g=1,2,3$, see s.\ref{mumm} below.

However, the set of modular forms is by no means exhausted
by these trivial characters of the permutation group.
Most important are other examples, having the same form
for all $g$, like
\be
\xi_{2+4k,2+4l}  \equiv
\sum_{e,e'}^{N_e} <e,e'>
\theta_{e}^{4+8k}\theta_{e'}^{4+8l}
= \sum_e^{N_e} \theta_e^{4+8k}\xi_{2+4l}[e]
\ee
which has weight $4(k+l+1)$.
Modular invariance of $\xi_{2+4k,2+4l}$ implies that
\be
\xi_{2+4l}[e] \equiv \sum_{e'}^{N_e} <e,e'>\theta_{e'}^{4+8l}
\ee
transforms under modular transformations exactly like $\xi_e^4$
(we call such forms "semi-modular").
The sign factors $<e,e'>$ serve to restore modular invariance
whenever $\theta^4_{e'}$ appear instead of $\theta_{e'}^8$.

As discovered in \cite{DHP1}-\cite{DHP9}, \cite{CP}-\cite{CPG2}
and formulated in a very clear and general form in \cite{Gru},
superstring measures are actually constructed from a wider
family of modular forms of weight $8$, of which $\xi_8$, and
$\xi_4^2$ and $\xi_{2,6}$ are just the first three members:
\be
\xi_8^{(p)} = \sum_e^{N_e} \xi_8^{(p)}[e]
\ee
where
\be
\xi_8^{(0)}[e] = \theta_e^{16},
\ \ \ \ {\rm i.e.}\ \ \xi_8^{(0)}=\xi_8,\nn \\
\xi^{(1)}_8[e] = 
\theta_e^8 \sum_{e_1}^{N_e} \theta^8_{e+e_1}
=  \theta_e^8 \xi_4,
\ \ \ \ {\rm i.e.}\ \ \xi_8^{(1)}=\xi_4^2,\nn \\
\xi^{(2)}_8[e] = 
\theta_e^4\sum_{e_1,e_2}^{N_e}
\theta^4_{e+e_1}\theta^4_{e+e_2}\theta^4_{e+e_1+e_2}, \nn \\
\xi^{(3)}_8[e] = 
\theta_e^2\sum_{e_1,e_2,e_3}^{N_e}
\theta^2_{e+e_1}\theta^2_{e+e_2}\theta^2_{e+e_3}
\theta^2_{e+e_1+e_2}\theta^2_{e+e_1+e_3}\theta^2_{e+e_2+e_3}
\theta^2_{e+e_1+e_2+e_3}, \nn \\
\ldots
\label{xi8p1}
\ee
and in general
\be
\xi_8^{(p)}[e] = \sum_{e_1,\ldots,e_p}^{N_e}\left\{\theta_e\cdot
\left(\prod_i^p \theta_{e+e_i}\right)\cdot
\left( \prod_{i<j}^p\theta_{e+e_i+e_j}\right)\cdot
\left(\prod_{i<j<k}^p \theta_{e+e_i+e_j+e_k}\right)\cdot \ldots
\cdot\theta_{e+e_1+\ldots+e_p}\right\}^{4/2^p}
\label{xi8p}
\ee
Characteristics are added as vectors.
Sign factors $<e,e'>$ are not seen in these formulas, because,
say, in $\xi^{(2)}_8$
$$
<e,e+e_1><e,e+e_2><e,e+e_1+e_2>\ =\ <e,e_1>^2<e,e_2>^2=1,
$$
while in $\xi^{(3)}_8$
$$
\sqrt{<e,e+e_1>}\sqrt{<e,e+e_2>}\sqrt{<e,e+e_3>}
\sqrt{<e,e+e_1+e_2>}\sqrt{<e,e+e_1+e_3>}
\sqrt{<e,e+e_2+e_3>}\cdot 
$$ $$
\cdot\sqrt{<e,e+e_1+e_2+e_3>}
= \sqrt{<e,e_1>^4<e,e_2>^4<e,e_3>^4} =
<e,e_1>^2<e,e_2>^2<e,e_3>^2 = 1
$$
and so on.
Many terms in the sums (\ref{xi8p1}) and (\ref{xi8p})
are actually vanishing, because contributing
characteristics are odd, for careful analysis of this
phenomenon in terms of isotropic spaces and Lagrange varieties
see \cite{CPG1}.
Only $\xi^{(p)}_8$ with $p\leq g$ appear in NSR measures in
s.\ref{NSRm} below.
For $g\geq 5$ fractional powers of theta-constants begin to appear
in the relevant $\xi^{(p)}_8$, what is by no means a drawback:
see \cite{Ma} for a very optimistic
analysis of the $g=5$ case.

\subsubsection{Grushevsky's basis}

In \cite{Gru} a slightly different basis was actually used,
with all diagonal terms eliminated from the sums (\ref{xi8p1})
and (\ref{xi8p}):
\be
\xi^{(0)}_8[e] = G^{(0)}_8[e], \nn \\
\xi^{(1)}_8[e] = G^{(0)}_8[e] + G^{(1)}_8[e], \nn \\
\xi^{(2)}_8[e] = G^{(0)}_8[e] + 3G^{(1)}_8[e] + G^{(2)}_8[e] , \nn \\
\xi^{(3)}_8[e] = G^{(0)}_8[e] + 7G^{(1)}_8[e] + 7G^{(2)}_8[e]
+ G^{(3)}_8[e], \nn \\
\xi^{(4)}_8[e] = G^{(0)}_8[e] + 15G^{(1)}_8[e] + 35G^{(2)}_8[e]
+ 15G^{(3)}_8[e] + G^{(4)}_8[e], \nn \\
\xi^{(5)}_8[e] = G^{(0)}_8[e] + 31G^{(1)}_8[e] + 155G^{(2)}_8[e]
+ 155G^{(3)}_8[e] + 31G^{(4)}_8[e] + G^{(5)}_8[e], \nn \\
\xi^{(6)}_8[e] = G^{(0)}_8[e] + 63G^{(1)}_8[e] + 651G^{(2)}_8[e]
+ 1395G^{(3)}_8[e] + 651G^{(4)}_8[e] + 63G^{(5)}_8[e]
+ G^{(6)}_8[e], \nn \\
\ldots
\label{Grub}
\ee
and in general
\be
\xi^{(p)}_8[e] = G^{(p)}_8[e] + (2^p-1)G^{(p-1)}_8[e] +
\frac{(2^p-1)(2^{p-1}-1)}{3}G^{(p-2)}_8[e]
+ \frac{(2^p-1)(2^{p-1}-1)(2^{p-2}-1)}{7\cdot 3}G^{(p-3)}_8[e] +
\nn \\
+ \frac{(2^p-1)(2^{p-1}-1)(2^{p-2}-1)(2^{p-3}-1)}{15\cdot 7\cdot 3}
G^{(p-4)}_8[e] +
\frac{(2^p-1)(2^{p-1}-1)(2^{p-2}-1)(2^{p-3}-1)(2^{p-4}-1)}
{31\cdot 15\cdot 7\cdot 3}G^{(p-5)}_8[e] + \ldots
\nn
\ee
(note the reversed order of terms in the last formula).
The definition of, say, $G_8^{(1)}$ is
\be
G_8^{(1)}[e] \equiv \theta_e^8 \sum_{e_1\neq 0}^{N_e} \theta^8_{e+e_1}
= \theta_e^8 \left(\sum_{e_1}^{N_e} \theta^8_{e+e_1} -
\theta_e^8\right) = \xi^{(1)}_8[e] - \xi^{(0)}_8[e]
\ee
In other words, in the sum for $\xi^{(1)}_8[e]$ there is
one term with $e_1=0$, which is $G^{(0)}_8$, and all the rest is
$G^{(1)}_8$.
Similarly, in the double sum for $\xi_8^{(2)}$ there is a
contribution from $e_1=e_2=0$ -- this is $G^{(0)}_8$,--
there are contributions from either $e_1=0$ and $e_2\neq 0$ or
$e_2=0$ and $e_1\neq 0$ or $e_1+e_2=0$ and $e_1=e_2\neq 0$ --
these are $3\cdot G^{(1)}_8$,-- and the rest is $G^{(2)}_8$.
When we proceed to triple sums, it is important to remember
that $e_1=e_2=0$ automatically implies that $e_1+e_2=0$:
this will produce factors like $2^p-4=4(2^{p-2}-1)$
instead of $2^p-3$
when we select the third characteristic to nullify after the
two are already chosen.
According to this definition $G^{(\!p\,)}=0$ for $p>g$.

There is no {\it a priori} reason to prefer $G^{(\!p\,)}_8$ over
$\xi^{(\!p\,)}_8$, but in \cite{Gru} it was demonstrated that
NSR measures are actually "more universal" (coefficients do
not depend on $g$) when expressed in
terms of $G^{(\!p\,)}_8$, see s.\ref{NSRm} below.

\subsubsection{Riemann identities}

There are no non-vanishing modular forms of weight $2$
made from the $4$-th powers of theta-constants, instead there
is a set of Riemann identities
\be
{\cal R}_* \equiv \sum_e^{N_e} <e,*>\theta_e^4 = 0
\label{Rid}
\ee
for all of the $N_*$ odd characteristics $*$.
Of $N_* = 2^{g-1}(2^g-1)$ Riemann identities there are
$\frac{1}{3}(4^g-1) = \frac{1}{3}(2^g+1)(2^g-1)$
linearly independent, and they reduce the number of
{\it linearly}-independent $\theta^4[e]$ from
$N_e = 2^{g-1}(2^g+1)$ to $\frac{1}{3}(2^g-1)(2^g+1)$.
Other relations between theta-constants involve
powers of $\theta^4$.
In naive superstring considerations an even stronger version
of Riemann identity is commonly used,
where up to three of the four theta-constants are promoted
to theta-functions:
\be
{\cal R}_*(\vec z_1,\vec z_2,\vec z_3|T) \equiv
\sum_e^{N_e} <e,*>\theta_e(\vec 0)\theta_e(\vec z_{12})
\theta_e(\vec z_{23})\theta_e(\vec z_{31}) = 0
\label{Ridz}
\ee
for any three vectors $\vec z_1$, $\vec z_2$, $\vec z_3$.
Both (\ref{Rid}) and (\ref{Ridz}) are corollaries of a
general relation
\be
\sum_{{\rm all}\ e}<e,*>\theta_e(\vec z_1)\theta_e(\vec z_2)
\theta_e(\vec z_3)\theta_e(\vec z_4) = \nn \\ =2^g
\theta_*\left(\frac{\vec z_1+\vec z_2+\vec z_3+\vec z_4}{2}\right)
\theta_*\left(\frac{\vec z_1+\vec z_2-\vec z_3-\vec z_4}{2}\right)
\theta_*\left(\frac{\vec z_1-\vec z_2+\vec z_3-\vec z_4}{2}\right)
\theta_*\left(\frac{\vec z_1-\vec z_2-\vec z_3+\vec z_4}{2}\right)
\ee
If one needs a sum over even characteristics at the l.h.s. it
is enough to add the same formula with
$\vec z_4 \rightarrow -\vec z_4$ to the r.h.s. (and divide by two).
In particular,
\be
\sum_e <e,*>\theta_e(\vec 0)^3\theta_e(\vec z) = 2^g
\theta_*^4\left(\frac{\vec z}{2}\right),
\ee
plays important role in superstring calculus.

\subsubsection{Decomposition rules}

For block-diagonal matrices $T = \left(\begin{array}{cc}
T_1 & 0 \\ 0 & T_2\end{array}\right)$ with $g=g_1+g_2$
the theta-functions factorize into products
$\theta_{e}(\vec z|T) = \theta_{e_1}(\vec z_1|T_1)
\theta_{e_2}(\vec z_2|T_2)$.
Above-mentioned modular forms behave as multiplicative characters
under this decomposition: they also factorize,
\be
\xi_{4k}(T) = \xi_{4k}(T_1)\xi_{4k}(T_2), \ \ \ \
\xi_{2+4k,2+4l}(T) = \xi_{2+4k,2+4l}(T_1)\xi_{2+4k,2+4l}(T_2), \nn \\
\xi^{(p)}_8[e](T) = \xi^{(p)}_8[e_1](T_1)\xi^{(p)}_8[e_2](T_2),
\ \ \ \ {\cal R}_*(T) = {\cal R}_{*_1}(T_1) {\cal R}_{*_2}(T_2),
\label{factomod}
\ee
while $\Pi$ in (\ref{Pif}) vanishes,
because some even characteristics $e$
get decomposed into two odd, for example
$\left[\begin{array}{cc} 1 & 1 \\ 1 & 1 \end{array}\right]
\rightarrow \left[\begin{array}{c} 1  \\ 1  \end{array}\right]
\otimes \left[\begin{array}{c} 1  \\ 1  \end{array}\right]$.

\subsection{Moduli space and Riemann $\theta$-functions
\cite{Mumb}-\cite{Ig}}

Riemann theta-functions are associated with tori which
are Jacobians of Riemann surfaces (complex curves).
Then $g$ is the genus of the curve and $T_{ij}$ is its
period matrix.
Period matrices define an embedding of moduli space of
Riemann surfaces into Siegel semi-space, and moduli
space has non-vanishing codimension $g(g+1)/2-(3g-3)$
for $g\geq 4$.
In terms of $T$ matrices this embedding is defined by a
set of transcendental {\it Shottky relations}.
Today the best known formulation of these relations is
that the corresponding theta-function is a $\tau$-function
of KP-hierarchy \cite{No}-\cite{KriSh} or, in other words, satisfy
the Wick theorem \cite{MPrev,UFN3,BMM}, 
\be
\det_{i,j} \frac{\theta_e(\vec x_i - \vec y_j)}
{E(x_i,y_j)\theta_e(\vec 0)}
= \frac{\theta_e(\sum_i \vec x_i - \sum_i \vec y_i)}
{\theta_e(\vec 0)}
\frac{\prod_{i<j} E(x_i,x_j) E(y_i,y_j)}{\prod_{i,j} E(x_i,y_j)}
\label{Wick}
\ee
generalizing a set of Gunning's, Fay's
\cite{Fay} and Welter's \cite{Welt} trisecant identities.
Here $E(x,y) = \frac{\theta_*(\vec x-\vec y)}{\nu_*(x)\nu_*(y)}$
is the prime form, $\vec x - \vec y = \int_y^x\vec\omega$
and $\nu^2(x) = \theta^*_{,i}(\vec 0)\omega_i(x)$.

Alternatively, one of the Shottky relations
(the only one in the case of $g=4$)
can be  formulated as the condition
\be
\chi_8 \equiv\ 2^g\xi_8 - \xi_4^2 \ =\
2^g\sum_e\theta_e^{16} - \Big(\sum_e \theta_e^8\Big)^2 = 0
\label{chif}
\ee
This is currently a hypothesis \cite{BK,BKMP,M1,DHP3},
rigorously proved only for $g=4$ \cite{Ig}
(for $g\leq 3$ this is not a Shottky relation, but
a simple algebraic relation in hyperelliptic parametrization,
see below).
At the same time it
expresses the equivalence (duality) of string compactifications
on $16$-dimensional tori with the two even self-dual lattices
$\Gamma_{16}$ and $\Gamma_8\times\Gamma_8$ and thus of
the heterotic $SO(32)$ and $E_8\times E_8$ strings \cite{heter}
and is strongly believed to be true "on physical grounds".

\subsection{Hyperelliptic surfaces \cite{Mumb,Fay,LM}}

Hyperelliptic surfaces are ramified double coverings
of Riemann sphere, which can be described as
\be
y^2 = \prod_{i=1}^{2g+2}(x-a_i)
\ee
Hyperelliptic surfaces form a $(2g-1)$-dimensional subspace
in the moduli space, parameterized by ramification points
$a_i$  modulo rational transformations
$(x,y|a_i) \rightarrow
\left(\frac{Ax+B}{Cx+D},\ \frac{y}{(Cx+D)^{g+1}}\Big|
\frac{Aa_i+B}{Ca_i+D}\right)$.
At genera $1$ and $2$ all Riemann surfaces are hyperelliptic.
At genus $3$ hyperelliptic locus has codimension $1$
and is defined by $\Pi=\prod_e \theta_e = 0$.

Consideration of hyperelliptic locus is very instructive,
because characteristic-dependence of theta-constants on it
becomes pure algebraic.
Semi-integer
theta-characteristics are associated with splitting of all
$2g+2$ ramification points into two sets of $g+1-2k$ and $g+1+2k$
points:
$\Big\{a\Big\} = \Big\{\tilde a\Big\}\bigcup
\Big\{\widetilde{\tilde a} \Big\}$.
Characteristic is even/odd if $k$ is even/odd,
it is also called singular if $k>2$. Non-vanishing are only
theta-constants associated with even non-singular characteristic,
$k=0$, and these non-vanishing
theta-constants are expressed through ramification points
by Thomae formulas:
\be
\theta^4[e] = \pm
(\det\sigma)^2\prod_{i<j}^{g+1}(\tilde a_i-\tilde a_j)
(\widetilde{\tilde a}_i-\widetilde{\tilde a}_j)
= \pm (\det\sigma)^2
\prod_{i<j}^{g+1} \tilde a_{ij}\widetilde{\tilde a}_{ij}
\label{Thom1}
\ee
Proportionality coefficient is transcendental, with
$\sigma_{ij} = \oint_{A_i} \frac{x^{j-1}dx}{y(x)}$,
see \cite{Mumb,Fay,LM} for details.
Fortunately, we do not need it in the present text.

In more detail Thomae formulas depend on the choice of
some set $U$ of $g+1$ ramification points.
Characteristics are in one-to-one correspondence with the
sets $S$, consisting of even numbers of ramification points.
Given $U$ and $S$ one can define a new set
$\ S\circ U = S\cup U - S\cap U\ $ and characteristic is
non-singular if $\#(S\circ U) = g+1$ and in this case
\be
\theta_{e}^4\ \sim\ (-)^{\#(S\cup U)}
\prod_{\stackrel{\tilde a_i \in S\circ U}
{\widetilde{\tilde a}_j \notin S\circ U}}
\left(\tilde a_i - \widetilde{\tilde a}_j\right)^{-1}
\label{Thom2}
\ee
The sign factor for any pair of characteristics (even or odd) is
\be
<e_1,e_2> = (-)^{\#(S_1\cup S_2)}
\label{sinS}
\ee

The number of non-singular even characteristics is
$N_{nse} = C^{g+1}_{2g+2}$, so that $N_{nse} = N_e$ for $g=1,2$,
while $N_{nse} = N_e-1$ for $g=3$ -- so that exactly one
even theta-constant vanishes and thus $\Pi = 0$
at codimension-one hyperelliptic locus in the moduli space at $g=3$.
The deviation from the hyperelliptic locus is measured by
$\sqrt{\Pi}$ which has modular weight $9$, and therefore the
relations between modular forms of lower weights (including those
of weight $8$, which are relevant for NSR measures) can
be exhaustively studied in hyperelliptic terms, i.e. pure
algebraically. To be more precise, if two forms of weight $\leq 8$
coincide at hyperelliptic locus at genus $3$, they coincide
everywhere.
At higher genera $g>3$ the codimension of hyperelliptic locus
in the moduli space is higher: $(3g-3)-(2g-1)=g-2$.
Of course, $\Pi=0$ at all these loci, but additional $g-3$ relations
occur which should also be taken into account, and also
Shottky relations should be added if one seeks for a description
in terms of modular forms.

On hyperelliptic locus the modular transformations
act by permutations of ramification points,
and modular forms are just symmetric polynomials of
$a_i$, multiplied by appropriate power of $\det\sigma$.
This makes hyperelliptic parametrization extremely convenient
for study of relations between modular forms, at least for
low genera and weights.

\subsection{Relations between modular forms at particular genera}

\subsubsection{Genus one}

Three theta-constants are related by Riemann identity
\be
\theta_{00}^4 = \theta_{01}^4 + \theta_{10}^4
\equiv b+c
\ee
The space of modular forms at genus one is generated by
two Eisenstein series:
\be
E_4 = \sum'_{m,n} \frac{1}{(m+n\tau)^4}
\sim \xi_4 = \sum_{e=1}^3 \theta_e^8
= (b+c)^2 + b^2 + c^2 = 2(b^2+bc+c^2)
\ee
and
\be
E_6 = \sum'_{m,n} \frac{1}{(m+n\tau)^6}
\sim \left(\theta\left[\begin{array}{c} 0\\1 \end{array}\right]^4-
\theta\left[\begin{array}{c} 1\\0 \end{array}\right]^4\right)
\left(\theta\left[\begin{array}{c} 0\\0 \end{array}\right]^4+
\theta\left[\begin{array}{c} 0\\1 \end{array}\right]^4\right)
\left(\theta\left[\begin{array}{c} 0\\0 \end{array}\right]^4+
\theta\left[\begin{array}{c} 1\\0 \end{array}\right]^4\right)
= \nn \\ = (b-c)(2b+c)(b+2c)
\ee
They are related to Dedekind function
$\eta = e^{i\pi\tau/12}\prod_{n=1}^\infty
\left(1 - e^{2\pi i n\tau}\right)$ by
\be
\eta^{24} = \Pi^8 = \Big(\theta_{00}\theta_{01}\theta_{10}\Big)^8 =
\Big(bc(b+c)\Big)^2 =
\frac{1}{1728}( E_4^3 - E_6^2 )
\ee
For any of the three even theta-characteristic $e$ we have:
\be
2\theta_e^{16} - \theta_e^8 \sum_{e'}^3 \theta_{e'}^8 =\
2<e,*>\theta_e^4 \prod^3_{e'} \theta_{e'}^4 \ \ \
=\ \ \ 2<e,*>\theta_e^4 \eta^{12}
= 2\theta_e^4\Pi_*^4
\label{chiRid}
\ee
i.e.
$$
2(b+c)^4 - (b+c)^2\cdot 2(b^2+bc+c^2) = 2(b+c) \cdot bc(b+c)
$$ $$
2b^4 - b^2\cdot 2(b^2+bc+c^2) = -2b \cdot bc(b+c)
$$ $$
2c^4 - c^2\cdot 2(b^2+bc+c^2) = -2c \cdot bc(b+c)
$$
Thus for $g=1$ the two vanishing-relations (\ref{Rid})
and (\ref{chif}) are actually the same.
Note that we absorbed the sign-factor $<e,*>$ into the definition
of $\Pi_*^4$.

Under modular transformations
$$
\begin{array}{|cccc|c|c|}
\hline &&&&&\\
&&&& \tau \rightarrow \tau + 1 & \tau \rightarrow -1/\tau \\
&&&&&\\
\hline
&&&&&\\
\theta_{00}^4 &=& b+c=& a & b & -a \\
\theta_{01}^4 &=&& b & a & -c \\
\theta_{10}^4 &=&& c & -c & -b \\
&&&&&\\
\hline
\end{array}
$$

For $g=1$ all our forms of weights $4$ and $8$ are expressed
through $\theta_e^8$, and $\xi_4=\sum_e \theta_e^8$:
\be
\xi_{2}[e] \equiv \sum_{e'}^3 <e,e'> \theta_{e'}^4
= 2\theta_e^4, \nn \\
\xi_{2,2} \equiv \sum_{e,e'}^3 \theta_e^4<e,e'> \theta_{e'}^4
= 2\sum_e^3\theta_e^8 = 2\xi_4, \nn \\
\xi_6[e] = \sum_{e'}^3 <e,e'> \theta_{e'}^{12} = -\theta_e^{12}
+ \frac{3}{2}\theta_e^4 \sum_{e'}^3\theta_{e'}^8
\ \stackrel{(\ref{chiRid})}{=} \
\theta_e^4 \sum_{e'}^3\theta_{e'}^8 - \Pi^4_* =
\xi_4\theta_e^4 - \Pi^4_*, \nn \\
\xi_{2,6} \equiv \sum_{e,e'}^3 \theta_e^4<e,e'>\theta_{e'}^{12}
= 2\sum_e^3\theta_e^{16} = 2\xi_8
\ \stackrel{(\ref{chif})}{=} \
\xi_4^2 =\left(\sum_e^3\theta_e^8\right)^2
\label{gen1theta12}
\ee
For the set of the CDG-Grushevsky forms
(\ref{xi8p1}) and (\ref{xi8p}) we have:
\be
\xi^{(\!p\,)}_8[e] = \alpha\!_p\,\theta_e^{16}
+ \beta_p\,\theta_e^8 \sum_{e'}^3 \theta_{e'}^8 =
\alpha\!_p\,\xi^{(0)}_8[e] + \beta_p\,\xi^{(1)}_8[e]
\ \stackrel{(\ref{chiRid})}{=}\
\frac{w\!_p}{2}\, \theta_e^8 \xi_4 +
\alpha\!_p\,\theta_e^{4}\Pi_*^4,
\label{gen1CDGG}
\ee
where $w_p = \alpha\!_p + 2\beta_p$. It follows that
\be
\xi^{(\!p\,)}_8 \equiv \sum_{e}^3 \xi^{(\!p\,)}_8[e] =
\frac{w_p}{2}\xi_4^2 = 2^{p-1}\xi_4^2
\label{xipxi4}
\ee
Numerical coefficients $\alpha\!_p$, $\beta_p$ and $w_p$ are easily
evaluated, if theta-constants are expressed through $b$ and $c$:


\be
g=1: \ \ \ \ \
\begin{array}{|c||c|c|c|}
\hline &&&\\
p & \alpha_p & \beta_p & w_p \\
&&&\\ \hline &&&\\
0 & 1 & 0 & 1 \\
1 & 0 & 1 & 2 \\
2 & -2 & 3 & 4 \\
3 & -6 & 7 & 8 \\
4 & -14 &15 & 16\\
\ldots &&&\\
p & -2(2^{p-1}-1) & 2^p-1 & \ \ 2^p\ \ \\
&&&\\
\hline
\end{array}
\label{wpvalues}
\ee


\noindent
In particular, it follows that
$\ \xi^{(2)}_8[e] = 2\theta_e^4\xi_6[e]$.

\bigskip

In hyperelliptic parametrization
\be
\theta_{00}^4 = a_{12}a_{34}, \ \ \ \theta_{01}^4 = a_{13}a_{24},
\ \ \ \theta_{10}^4 = a_{41}a_{23}
\label{gen1theta}
\ee
and formulas look a little more involved
than in terms of $b$ and $c$, for example:
\be
\xi_4 = \sum_e \theta_e^8 =
a_{12}^2a_{34}^2 + a_{13}^2a_{24}^2 + a_{14}^2a_{23}^2 =
-6s_4+6s_3s_1+\frac{7}{2}s_2^2-4s_2s_1^2+\frac{1}{2}s_1^4,
\ee
where $s_m = \sum_{k=1}^4 a_i^k$.
Also,
$$
\xi_8 = \sum_e \theta_e^{16} =
a_{12}^4a_{34}^4 + a_{13}^4a_{24}^4 + a_{14}^4a_{23}^4
= 2\xi_4^2
$$ $$
{\cal R}_* = \sum_e<e,*>\theta_e^4 \sim a_{12}a_{34}
- a_{13}a_{24} - a_{41}a_{23} = 0
$$ $$
\xi_{2,2} = a_{12}a_{34}
(a_{12}a_{34} + a_{13}a_{24} + a_{41}a_{23})
+ a_{13}a_{24}(a_{12}a_{34} + a_{13}a_{24} - a_{41}a_{23})
+a_{41}a_{23}(a_{12}a_{34} - a_{13}a_{24} + a_{41}a_{23})
$$
and
$$
\xi_{2,6} = a_{12}a_{34}(a_{12}^3a_{34}^3 + a_{13}^3a_{24}^3
+ a_{41}^3a_{23}^3)
+ a_{13}a_{24}(a_{12}^3a_{34}^3 + a_{13}^3a_{24}^3
- a_{41}^3a_{23}^3)
+a_{41}a_{23}(a_{12}^3a_{34}^3 - a_{13}^3a_{24}^3
+ a_{41}^3a_{23}^3)
$$
Still, all the relations, including (\ref{gen1CDGG}),
can be easily derived in this parametrization, and such
derivations are straightforwardly generalized to $g=2,3$.
The more economic $b,c$ parametrization is also generalizable
(it is related to expressions through theta-constants of
doubled argument, $\theta(2T)$, which was actually used in
\cite{CPG1}), but this is a slightly more involved technique,
unnecessary for our presentation.

Formula (\ref{Thom2}) looks as follows:

$$
\begin{array}{cccc|c}
S & S\cup U & S\cap U & S\circ U & \theta_e^4 \\
\hline &&&&\\
\emptyset & 34 & \emptyset & 34 &
\sim +\frac{1}{a_{31}a_{32}a_{41}a_{42}} \sim +a_{12}a_{34}\\
&&&&\\
13 & 134 & 3 & 14 &
\sim -\frac{1}{a_{12}a_{13}a_{42}a_{43}} \sim -a_{14}a_{23} \\
&&&&\\
14 & 134 & 4 & 13 &
\sim -\frac{1}{a_{12}a_{14}a_{32}a_{34}} \sim +a_{13}a_{24} \\
&&&&\\
23 & 234 & 3 & 24 &
\sim -\frac{1}{a_{21}a_{23}a_{41}a_{43}} \sim +a_{13}a_{24} \\
&&&&\\
24 & 234 & 4 & 23 &
\sim -\frac{1}{a_{21}a_{24}a_{31}a_{34}} \sim -a_{14}a_{23} \\
&&&&\\
1234 & 1234 & 34 & 12 &
\sim +\frac{1}{a_{13}a_{14}a_{23}a_{24}} \sim +a_{12}a_{34} \\
&&&&\\ \hline &&&&\\
12 & 1234 & \emptyset & 1234 & 0 \\
&&&&\\
34 & 34 & 34 & \emptyset & 0
\end{array}
$$

\bigskip
\noindent
It is assumed here that $U = \{a_3,a_4\}$:
this is the choice which reproduces (\ref{gen1theta}).
In the last two lines $\#(S\circ U)\neq g+1=2$, such sets $S$
correspond to the odd characteristic with vanishing theta-constant.

\subsubsection{Genus two}

Of six (as many as there are odd characteristics *)
Riemann identities (\ref{Rid}) there are five linearly independent,
and they express $10$ a priori different $\theta_e^4$
through $5$ linearly independent ones.
In addition there is one non-linear relation:
\be
\chi_8 = 4\xi_8 - \xi_4^2 = 0, \ \ \ \ \ \ \
{\rm i.e.}\ \ \ \ \xi^{(0)}_8 \equiv \xi_8 = \frac{1}{4}\xi_4^2,
\ \ \ \ \  \xi^{(1)}_8 = \xi_4^2
\ee
Further,
\be
\xi_{2,2} = 4\xi_4, \nn \\
\xi_{2,6} = 4\xi_8 = \xi_4^2
\label{gen2xi26}
\ee
and
\be
\xi_8^{(2)}[e] = 4\theta_e^4 \xi_{6}[e],  \ \ \ \ \
\xi_8^{(2)}= \sum_{e}^{10} \xi_8^{(2)}[e] = 4\xi_{2,6} = 4\xi_4^2
\ee
\be
\xi^{(\!p\,)}_8[e] = \alpha\!_p\,\theta_e^{16}
+ \beta_p\,\theta_e^8 \sum_{e'}^3 \theta_{e'}^8
+ \gamma_p\,\theta_e^4 \sum_{e',e''}^3 \theta_{e'}^4\theta_{e''}^4
\theta^4_{e+e'+e''} =
\alpha\!_p\,\xi^{(0)}_8[e] + \beta_p\,\xi^{(1)}_8[e] +
\gamma_p\,\xi^{(2)}_8[e]
\label{gen2CDGG}
\ee
It follows that
\be
\xi^{(\!p\,)}_8 \equiv \sum_{e}^3 \xi^{(\!p\,)}_8[e] =
\left(\frac{1}{4}\alpha_p + \beta_p + 4\gamma_p\right)\xi_4^2
= \frac{1}{4}\,w_p\,\xi_4^2
\ee
where $w_p = \alpha\!_p + 4\beta_p + 16\gamma_p$.
Numerical coefficients $\alpha\!_p$, $\beta_p$ and $\gamma_p$
are easily evaluated if theta-constants are expressed in
hyperelliptic parametrization, where they become simple
algebraic relations.

\be
g=2: \ \ \ \ \
\begin{array}{|c||c|c|c||c|}
\hline &&&&\\
p & \alpha_p & \beta_p & \gamma_p & w_p \\
&&&&\\ \hline &&&&\\
0 & 1 & 0 & 0 & 1 \\
1 & 0 & 1 & 0 & 4 \\
2 & 0 & 0 & 1 & 16 \\
3 & 8 & -14 & 7 & 64 \\
4 & 56 & -90 & 35 & \ \ \ 256\ \ \\
\ldots &&&&\\
p & \frac{8(2^{p-1}-1)(2^{p-2}-1)}{3} & -2(2^p-1)(2^{p-2}-1)
&\frac{(2^p-1)(2^{p-1}-1)}{3} & 4^p \\
&&&&\\
\hline
\end{array}
\ee

\bigskip

\noindent
The simplest way to prove this kind of identities is to use
hyperelliptic parametrization, where they become simple
algebraic relations.
In the basis selected in \cite{CP} -- it corresponds to
taking $U=\{a_2,a_3,a_5\}$ in (\ref{Thom2})\footnote{
However, association of theta-characteristics -- the map
$S\rightarrow e(S)$ -- in \cite{CP} does not look consistent
with the rule (\ref{sinS}), and we choose another one in the
second line of the table.} -- we have:
$$
{\bf odd\ characteristics:}
\ \ \ \ \ \
\begin{array}{|c|cccccc|}
\hline &&&&&& \\
S & 14 & 16 & 46 & 23 & 25 & 35 \\
& 2356 & 2345 & 1235 & 1456 & 1346 & 1246 \\
&&&&&& \\\hline &&&&&& \\
e(S)
&\left[\begin{array}{cc} 0&1\\0&1\end{array}\right]
&\left[\begin{array}{cc} 1&0\\1&1\end{array}\right]
&\left[\begin{array}{cc} 1&1\\1&0\end{array}\right]
&\left[\begin{array}{cc} 0&1\\1&1\end{array}\right]
&\left[\begin{array}{cc} 1&1\\0&1\end{array}\right]
&\left[\begin{array}{cc} 1&0\\1&0\end{array}\right] \\
&&&&&& \\\hline &&&&&& \\
\theta_e&0&0&0&0&0&0 \\
&&&&&& \\\hline
\end{array}
$$

{\bf even characteristics:}

\bigskip

\centerline{ {\footnotesize
$
\begin{array}{|c|cccccccccc|}
\hline &&&&&&&&&& \\
S &\emptyset &24&13&56&26&45&15&36&34&12 \\
&123456 &1356&2456&1234&1345&1236&2346&1245&1256&3456 \\
&&&&&&&&&& \\ \hline &&&&&&&&&& \\
e(S) & \left[\begin{array}{cc} 0&0\\0&0\end{array}\right]
& \left[\begin{array}{cc} 0&0\\0&1\end{array}\right]
& \left[\begin{array}{cc} 0&0\\1&1\end{array}\right]
& \left[\begin{array}{cc} 0&0\\1&0\end{array}\right]
& \left[\begin{array}{cc} 1&1\\1&1\end{array}\right]
& \left[\begin{array}{cc} 1&1\\0&0\end{array}\right]
& \left[\begin{array}{cc} 1&0\\0&1\end{array}\right]
& \left[\begin{array}{cc} 1&0\\0&0\end{array}\right]
& \left[\begin{array}{cc} 0&1\\1&0\end{array}\right]
& \left[\begin{array}{cc} 0&1\\0&0\end{array}\right] \\
&&&&&&&&&&\\ \hline &&&&&&&&&& \\
\theta^4_e &-a_{146}a_{235} & a_{126}a_{345} & a_{125}a_{346}
&-a_{145}a_{236}&a_{124}a_{356}&-a_{156}a_{234}&a_{123}a_{456}
&-a_{134}a_{256}&-a_{136}a_{245}&-a_{135}a_{246}\\
&&&&&&&&&&\\
\hline
\end{array}
$
}}

\bigskip

Note that there is no direct counterpart of the
relation (\ref{chiRid}) already for $g=2$:
the form $\chi_8 = 4\xi_8-\xi_4^2$ is not a linear
combination of Riemann identities (\ref{Rid}).
Moreover, one can easily check that it does not
automatically vanish for arbitrary
set of $5$ linearly-independent $\theta_e^4$:
from genus two
$\chi_8=0$ is an {\it additional} relation between
theta-constants, algebraically (not only linear)
independent of Riemann identities.

\subsubsection{Genus three}

The number $N_*$ of Riemann identities is now $28$,
of which $\frac{4^g-1}{3} = 21$ are linearly independent
and there are $\frac{(2^g+1)(2^{g-1}+1)}{3}=
36-21 = 15$ linearly independent $\theta_e^4$.
Again, there are additional non-linear relations,
including
\be
\chi_8= 8\xi_8 - \xi_4^2 =\
8\sum_{e}^{36} \theta_e^{16} -
\left(\sum_e^{36}\theta_e^8\right)^2 = 0
\ee
Hyperelliptic locus has codimension one in moduli space
and is defined by $\Pi = \prod^{36}_{e} \theta_e = 0$.
Still, hyperelliptic parametrization can be used to
prove formulas at genus $3$ for modular functions of weights
$\leq 8$,  because deviations from hyperellipticity are
proportional to $\sqrt{\Pi}$ which has weight $9$.

\subsubsection{Genus four}

As shown in \cite{Ig}, and widely used since \cite{BK,M1,DHP1},
$\chi_8 = 0$ exactly at the moduli space, embedded as
codimension-one subspace in the Siegel upper semi-space.
Hyperelliptic locus now has codimension $g-2=2$, this is
the place where $\Pi=0$, but actually not just one, but $10$
out of $136$ even  theta-constants vanish on it (though it is not 
the only place in the Siegel half-space where such things happen).
Simple hyperelliptic calculations are still very useful here,
but are not as conclusive as they are for $g<4$.

\section{Mumford measure for critical bosonic string
\cite{BKMP,M1} \label{mumm}}

After a brief exposition of the theory of theta-constants
-- note that we do not need anything more than above simple
statements -- we are ready to switch to the string measures.
As already mentioned in the Introduction, Belavin-Knizhnik
theorem \cite{BK} expresses them through the holomorphic
Mumford measure on the moduli space of complex curves,
which has degree-2 poles at the boundaries:
namely when one of the cycles (contractible or
non-contractible) gets shrinked.
The degree of the pole is controlled by the negative mass
squared of a tachyon, present in the spectrum of bosonic string.
Residues at the poles are given by two-point a function
in the case of non-contractible cycle (when genus $g$
curve degenerates into the one of $g-1$) and a product
of two one-point functions in the case of contractible cycle
(when the curve splits into two of genera $g_1$ and $g_2=g-g_1$).
In fact the values of pole degrees are enough to determine
the measure and above properties can be used to read off
expressions for one- and two-point functions.
The most interesting object is the string measure on the
{\it universal moduli space}, unifying all genera and all
the correlators (scattering amplitudes) \cite{UMS}.
$n$-point correlators can also be promoted to stringy correlators
by inclusion of Riemann surfaces with boundaries and/or
non-oriented \cite{open}.

In fact all these generalizations are rather straightforward once
the structure of string measures for particular genera is
clarified\footnote{The only subject which remains really puzzling
concerns {\it arithmetic} properties of Mumford measure
\cite{DJS,LevM}. Especially interesting is the relation between
Polyakov and Migdal formalisms for string measures: the latter
one is based on the use of equilateral triangulations, i.e.
rational surfaces (Grothendieck's dessins d'enfant), which are
not very well distributed inside the moduli space what makes
equivalence of measures a kind of surprise, see \cite{LevM}
for details.}
-- and we list here original expressions from \cite{BKMP,M1}.
For somewhat less explicit expressions for all genera see
\cite{VV}-\cite{MPrev}.

{\bf Genus one:}
\be
\frac{1}{({\rm Im}\ \tau)^{14}}\left|\frac{d\tau}
{\left(\prod_{e}^3\theta[e](\tau)\right)^8}\right|^{2}
\ \ \ \ \ \ \ {\rm i.e.} \ \ \ \ \ d\mu = \frac{d\tau}{\Pi^8}
\label{gen1bos}
\ee

{\bf Genus two:}
\be
\frac{1}{\big(\det\,({\rm Im}\  T)\big)^{13}}\left|
\frac{dT_{11} dT_{12} dT_{22}}
{\left(\prod_{e}^{10}\theta[e](\tau)\right)^2}\right|^{2}
\ \ \ \ \ \ \ {\rm i.e.} \ \ \ \ \
d\mu = \frac{\prod_{i<j}^2 dT_{ij}}{\Pi^2}
\ee

{\bf Genus three:}
\be
\frac{1}{\big(\det\,({\rm Im}\  T)\big)^{13}}\left|
\frac{dT_{11} dT_{12} dT_{13} dT_{22} dT_{23} dT_{33}}
{\left(\prod_{e}^{36}\theta[e](\tau)\right)^{1/2}}\right|^{2}
\ \ \ \ \ \ \ {\rm i.e.} \ \ \ \ \
d\mu = \frac{\prod_{i<j}^3 dT_{ij}}
{\sqrt{\phantom{5^{5^5}}\!\!\!\!\!\!\!\!\Pi}}
\ee
Zero of the form in denominator is at the hyperelliptic locus.
The square root singularity at this locus is fictitious:
the period matrix in the vicinity of the locus is a {\it square}
of the proper modulus \cite{BKMP,M1}.

{\bf Genus four:} \\
This is the first time when the module space is smaller then
Teichmuller one, it has complex codimension one and
is defined by the zero of a single Shottky condition
\be
\chi_8 = 0
\ee
where $\chi_8$ is the weight-8 modular form on Teichmuller
space,
\be
\chi_8(T) =
16\sum_e \theta[e]^{16} - \left(\sum_e\theta[e]^8\right)^2
\ee
Bosonic string  measure is
\be
\frac{1}{\big(\det\,({\rm Im}\  T)\big)^{13}}\left|
\frac{\prod_{ij\leq j}^4 dT_{ij}}{\chi_8(T)}\right|^{2}
\label{gen4bosm}
\ee
This wonderful formula, suggested in \cite{BK} and \cite{M1}
never attracted attention that it deserves and was not
investigated as carefully as its lower-genera counterparts.
Note that instead of the holomorphic delta-function of $\chi_8$
in (\ref{gen4bosm}) one can put the sum of the NSR measures
$\sum_e \Xi_8[e]$, which vanishes on the moduli space and is
essentially the same as $\chi_8$.

\section{NSR measures}

\subsection{Superstring from NSR measures for fermionic string}

Superstring possesses space-time supersymmetry in critical
dimension $d=10$.
Two approaches are developed in order to describe it in
the first quantization formalism, i.e. with the help of
the two-dimensional actions on string world sheet.
One approach (Green-Schwarz formalism \cite{GSst}-\cite{Berk})
is explicitly $d=10$ supersymmetric, but the
two-dimensional action is highly non-linear and possesses
sophisticated $\kappa$-symmetry.
Another, NSR approach \cite{superst,GSO}
is based on the theory of {\it fermionic string},
defined as possessing the world-sheet, i.e. $2d$ supersymmetry.
On world sheets with non-trivial topologies one can
impose a variety of boundary conditions on $2d$ fermions,
associated with different spin-structures or, what is the same,
the theta-characteristics.
The corresponding holomorphic NSR measures $d\mu[e]$ on the
moduli space of Riemann surfaces also
depend on theta-characteristics.
Fermionic string does not have $10d$ space-time supersymmetry,
it has tachyon and divergencies, just as bosonic string.
However, superstring Hilbert space is just a subspace in
the Hilbert space of fermionic space, and the relevant GSO
projection \cite{GSO} is provided simply by a sum of any
holomorphic conformal block over the spin-structures:
\be
\Big< A \Big> = \int \frac{1}{\Big(\det\, ({\rm Im}\ T)\Big)^5}
\left|\sum_e A[e]d\mu_[e]\right|^2
\ee
where $A[e]$ is a combination of holomorphic Green functions,
associated with the multi-point observable $A$.

In genus one the three NSR measures are well known \cite{superst}:
\be
d\mu[e] = \frac{<e,*>\theta_e^4\ d\tau}{\eta^{12}},
\ee
what means that they are expressed through Mumford measure
$d\mu = \frac{d\tau}{\eta^{24}} = \frac{d\tau}{\Pi^8}$
from (\ref{gen1bos}):
\be
d\mu_e = <e,*>\theta_e^4 \eta^{12} d\mu = \theta_e^4 \Pi_*^4 d\mu
\label{gen1NSRm}
\ee
where $*$ is the only odd theta-characteristic at $g=1$.
(Of course, for genus one the measure includes the $6$-th power
of $\ {\rm Im}\ \tau\ $ instead of the $5$-th one in for $g>1$.)

It is an old conjecture that the situation is similar
for arbitrary genus:
\be
d\mu[e] = \Xi_8[e] d\mu,
\label{NSRconj1}
\ee
where $\Xi_8[e]$ is a semi-modular form of weight $8$.
This is a non-trivial hypothesis for $g\geq 4$, because
there is no obvious reason why $d\mu[e]/d\mu$ should have
any nice continuation to entire Siegel space, beyond the
moduli space.
Still, if this hypothesis is true, for any correlator
in superstring theory we have a simple representation in
terms of an integral over moduli space:
\be
\Big< A \Big> = \int \frac{|d\mu|^2}
{\Big(\det\, ({\rm Im}\ T)\Big)^5}
\left|\sum_e A[e]\, \Xi_8[e]\,\right|^2
\ee
Under these assumptions the only unknown is the set of
forms $\Xi_8[e]$, which should satisfy two simple
properties: factorization and the condition of vanishing
cosmological constant,
\be
\sum_e d\mu[e] = 0, \ \ \ \ \ {\rm i.e.}\ \ \ \ \
\sum_e \Xi_8[e] = 0
\label{vacoc}
\ee
For genus $1$ eq.(\ref{vacoc}) for (\ref{gen1NSRm})
is an immediate corollary of the Riemann identity (\ref{Rid}),
\be
\sum_e <e,*> \theta[e]^4 = 0
\label{Rid1}
\ee
It seemed a natural generalization of conjecture (\ref{NSRconj1})
to extend this property to all genera \cite{MP1,AS}:
\be
\Xi_8[e]\ \stackrel{?}{=}\  <e,*>\theta_e^4K^*_6,
\label{NSRconj2}
\ee
especially because (\ref{Ridz}) would then automatically guarantee
the vanishing of all $g\geq 1$ corrections to the $1,2,3$-point
functions.
Immediate drawback of this Riemann-identity hypothesis was
explicit dependence on the odd characteristic $*$, which would
un-acceptedly show up in non-vanishing $4$-point function and
in higher correlators.
Worse than that, an appropriate form $K_6^*$ does not seem
to exist.

It was believed that the NSR measure can be {\it derived},
starting from explicitly $2d$-supersymmetric formalism for
fermionic string, based on the clever definition of
super-Riemann surfaces, by integrating over odd supermoduli.
However, naive simplified approaches of this kind (attempting
to trivialize the supermoduli bundle over the ordinary module
space) failed, and accurate integration was performed only
recently in \cite{DHP1}-\cite{DHP4} and only for $g=2$.
The outcome was a confirmation of hypothesis
(\ref{NSRconj1}) and a clear denunciation of (\ref{NSRconj2}):
it appeared that instead of continuing (\ref{Rid1}) from
$g=1$ to $g>1$ one should rather substitute it by
\be
{\bf g=1}: \ \ \ \ \ \
\Xi_8[e]=
\sum_e <e,*> \theta[e]^4 \Pi_*^4 \ \stackrel{(\ref{chiRid})}{=}\
2\sum_e \theta_e^{16} - \left(\sum_e \theta_e^8\right)^2 =
\chi_8 \ \stackrel{(\ref{xi8p1})}{=}\ 2\xi^{(0)}_8-\xi^{(1)}_8
\label{chiRid1}
\ee
and continue the r.h.s. (note that relation (\ref{chiRid})
does not survive at $g\geq 2$, so that continuations of its
two sides deviate from each other).
Such continuation was {\it derived} in \cite{DHP1}-\cite{DHP4}
for $g=2$, reformulated and generalized to $g=3,4$ in
\cite{CP,CPG1,CPG2} and was put in the nice form, conjecturally
reasonable for arbitrary $g$ in \cite{Gru}.
Since CPG-Grushevsky conjecture for $g\geq 3$ expresses
$d\mu[e]$ through $\xi^{(p)}_8$ with $p\geq 3$, it does not
contain an explicit $\theta_e^4$ factor, what makes puzzling
the story about the $1,2,3$-point functions.

\subsection{Anzatz for the NSR measures \cite{DHP1,CPG1,Gru,CPG2}
\label{NSRm}}

The natural generalization of the r.h.s. of (\ref{chiRid1}) is
\be
{\bf any\ g:}\ \ \ \ \ \ \
\Xi_8[e] = \sum_{p=0}^g h_p\,\xi^{(p)}_8[e],
\label{xior}
\ee
where CDG-Grushevsky forms at the r.h.s. are defined in
(\ref{xi8p1}) and (\ref{xi8p}) and coefficients $h_p$
are constrained by requirements of factorization and vanishing
of the cosmological constant.

The latter one implies that
\be
\sum_{e}^{N_e} \Xi_8[e] = \sum_{p=0}^g h_p\,\xi^{(p)}_8 = 0
\label{vaco1}
\ee
Since the l.h.s. is a modular form of weight $8$, it has good chances
to be proportional to $\xi_4^2 \ \stackrel{(\ref{chif})}{=} 2^g\xi_8$
and the same is actually true for all the terms in the sum:
\be
\xi^{(p)}_8 = \frac{1}{2}W_{\!p}\,\xi_4^2
\label{Wpdef}
\ee
Thus the requirement (\ref{vaco1}) simply states that
\be
\sum_{p=0}^g h_pW_p = 0
\label{vaco2}
\ee
Coefficients $W_p$ can be evaluated by different methods,
but the simplest one is to go to the high-codimension subset
at the boundary of moduli space, when the curve degenerates
into a set of tori and period matrix $T$ becomes diagonal
$T = {\rm diag}(\tau_1,\ldots,\tau_g)$.
Then $\xi_4(T) \rightarrow \prod_{i=1}^g\xi_4(\tau_i)
= \xi_4^{\otimes g}$ and
\be
\xi^{(p)}_8(T) \longrightarrow \prod_{i=1}^g \xi^{(p)}_8(\tau_i)
\ \stackrel{(\ref{xipxi4})}{=}\
\left(\frac{w_p}{2}\right)^g \prod_{i=1}^g\xi_4^2(\tau_i)
\label{factogen1}
\ee
so that
\be
W_p = 2\left(\frac{w_p}{2}\right)^g
\ \stackrel{(\ref{wpvalues})}{=}\ 2^{g(p-1)+1}
\ee

\bigskip

Of course, (\ref{vaco2}) is an important but non-restrictive
constraint on the coefficients $h_p$.
All the $h_p$ are determined if the same reduction to genus one
is made for the individual $\Xi_8[e]$:
On one side,
\be
\Xi_8[e](T) \rightarrow \prod_{i=1}^g \Xi_8[e_i](\tau_i)
\ \stackrel{(\ref{gen1NSRm})}{=}\
\prod_{i=1}^g \Big\{\theta_{e_i}^4\Pi^4_*(\tau_i)\Big\}
\ee
on another side
\be
\Xi_8[e](T) \ \stackrel{(\ref{vaco1})}{=}\
\sum_{p=0}^g h_p\,\xi^{(p)}_8[e] \longrightarrow
\sum_{p=0}^g h_p
\left\{\prod_{i=1}^g \xi^{(p)}_8[e_i](\tau_i)\right\}
\ \stackrel{(\ref{gen1CDGG})}{=}\
\sum_{p=0}^g h_p\left\{
\prod_{i=1}^g \left(\frac{w_p}{2}\theta^8_{e_i}\xi_4
+ \alpha\!_p\,\theta_{e_i}^4\Pi^4_*\right)(\tau_i)\right\}
\ee
Comparing the two expressions we obtain a set of $g+1$
linear equations for $g+1$ coefficients $h_p$:
\be
\sum_{p=0}^g h_p w_p^k (2\alpha_p)^{g-k} = 2^g\delta_{k,0}
\ \ \ \ \ {\rm or} \ \ \ \ \
\sum_{p=0}^g \tilde h\!_p\,\lambda_p^k = 2^g\delta_{k,0}
\ee
with $k=0,\ldots,g$,
$\ \ \tilde h_p = (2\alpha_p)^g\tilde h_p$ and
$\ \lambda_p = w_p/2\alpha_p$,
so that
$h_p$ is the ratio of Van-der-Monde determinants:
\be
\tilde h\!_p = 2^g\frac{\Delta_p(\lambda)}{\Delta(\lambda)}
= 2^g\prod_{i\neq p}^g \frac{\lambda_i}{\lambda_i-\lambda_p}
\ \ \ \ {\rm and} \ \ \ \
h\!_p = \prod_{i\neq p}^g \frac{w_i}{w_i\alpha_p-w\!_p\alpha_i}
\ee

\bigskip

\be
\begin{array}{|c|ccccccc|}
g & h_0 & h_1 & h_2 & h_3 & h_4 & h_5 & \ldots \\
\hline &&&&&&&\\
1 &1&-\frac{1}{2}&&&&&\\&&&&&&&\\
2 &\frac{2}{3}&-\frac{1}{2}&\frac{1}{12}&&&&\\&&&&&&&\\
3 &\frac{8}{21}&-\frac{1}{3}&\frac{1}{12}&-\frac{1}{168}&&&\\&&&&&&&\\
4 &\frac{64}{315}&-\frac{4}{21}&\frac{1}{18}&-\frac{1}{168}
&\frac{1}{5040}&&\\&&&&&&&\\
5 &\frac{1024}{9765}&-\frac{32}{315}&\frac{2}{63}&-\frac{1}{252}
&\frac{1}{5040}&-\frac{1}{312480}&\\&&&&&&&\\
\ldots &&&&&&&\\
\end{array}
\label{hps}
\ee

\bigskip
\noindent
It is easy to check, that
the vanishing relations (\ref{vaco2}) and thus (\ref{vaco1})
are true with these values of $h\!_p$.

In Grushevsky's basis \cite{Gru} the coefficients are much nicer,
moreover, they are actually independent of $g$.
Indeed, substituting $\xi^{(p)}_8$ in the form (\ref{Grub})
and $h_p$ from the table into (\ref{xior}) we obtain:
\be
\begin{array}{c|cccccccccccc}
g=1 & \Xi_8[e] = \frac{1}{2}\Big(G^{0}_8[e] &-& G^{(1)}_8[e]\Big)
&& && && && \\
g=2 & \Xi_8[e] = \frac{1}{4}\Big(G^{0}_8[e] &-& G^{(1)}_8[e]
&+& \frac{1}{3} G^{(2)}_8[e] \Big) && && && \\
g=3 & \Xi_8[e] = \frac{1}{8}\Big(G^{0}_8[e] &-& G^{(1)}_8[e]
&+& \frac{1}{3} G^{(2)}_8[e] &-& \frac{1}{21} G^{(3)}_8[e]
\Big) && && \\
g=4 & \Xi_8[e] = \frac{1}{16}\Big(G^{0}_8[e] &-& G^{(1)}_8[e]
&+& \frac{1}{3} G^{(2)}_8[e]&-& \frac{1}{21} G^{(3)}_8[e]
&+& \frac{1}{315} G^{(4)}_8[e]\Big) &&   \\
&& && && && &&&\\ \hline && && && && &&&\\
g=5 & \Xi_8[e] = \frac{1}{32}\Big(G^{0}_8[e] &-& G^{(1)}_8[e]
&+& \frac{1}{3} G^{(2)}_8[e]&-& \frac{1}{21} G^{(3)}_8[e]
&+& \frac{1}{315} G^{(4)}_8[e] &-& \frac{1}{9765} G^{(5)}_8[e]\Big)
\\
\ldots
\end{array}
\nn
\ee
and finally
\be
d\mu[e] = \Xi_8[e] d\mu, \ \ \ \ \ \ \ \
\Xi_8[e] = \frac{1}{2^g}\sum_{p=0}^g
\frac{(-)^p}{\prod_{i=1}^p(2^i-1)}\,
G^{(p)}_8[e]
\label{NSRGru}
\ee
(the coefficient in the term with $p=0$ is unity, by the
usual rule $\prod_1^0 = 1$, like $0!=1$).
Note that in \cite{Gru} the normalization of $G^{(p)}_8$
was chosen differently, therefore the coefficients
in (\ref{NSRGru}) are also different.

\subsection{More degeneration examples}

In addition to (\ref{factogen1})
one can consider reductions to lower-codimension
components of the boundary, where, for example, the curve
degenerates into two of genera $g_1$ and $g_2$ with
$g_1+g_2=g$. This is an important check, but the result
actually follows from above much simpler consideration.

For example, the genus-three
\be
\Xi_8 \ \stackrel{(\ref{xior})}{=}\
\frac{8}{21}\xi_8^{(0)} - \frac{1}{3}\xi_8^{(1)}
+ \frac{1}{12}\xi_8^{(2)} - \frac{1}{168}\xi_8^{(3)}
\ee
decomposes into genus-one and genus-two quantities
\be
\Xi_8 \longrightarrow \Xi_8
\left(\begin{array}{ccc} \tau & 0 & 0 \\
0 & T_{11} & T_{12}\\0 & T_{12} & T_{22}
\end{array}\right) =
\frac{8}{21}\xi_8^{(0)}(\tau)\otimes \xi_8^{(0)}
\left(\begin{array}{cc} T_{11} & T_{12}\\T_{12} & T_{22}
\end{array}\right)
- \frac{1}{3}\xi_8^{(1)}(\tau)\otimes \xi_8^{(1)}
\left(\begin{array}{cc} T_{11} & T_{12}\\T_{12} & T_{22}
\end{array}\right) + \nn \\
+ \frac{1}{12}\xi_8^{(2)}(\tau)\otimes \xi_8^{(2)}
\left(\begin{array}{cc} T_{11} & T_{12}\\T_{12} & T_{22}
\end{array}\right)
- \frac{1}{168}\xi_8^{(3)}(\tau)\otimes \xi_8^{(3)}
\left(\begin{array}{cc} T_{11} & T_{12}\\T_{12} & T_{22}
\end{array}\right) = \ \ \ \ \ \ \
\nn \\  \ \stackrel{(\ref{gen12CDGG})}{=} \
 \left(\xi^{(0)}_8 - \frac{1}{2}\xi^{(1)}_8\right)(\tau)
\otimes
\left(\frac{2}{3}\xi^{(0)}_8 - \frac{1}{2}\xi^{(1)}_8
+ \frac{1}{12}\xi^{(2)}_8\right)
\left(\begin{array}{cc} T_{11} & T_{12}\\T_{12} & T_{22}
\end{array}\right)  \ \stackrel{(\ref{xior})}{=}\ \
\Xi_8(\tau)\otimes \Xi_8
\left(\begin{array}{cc} T_{11} & T_{12}\\T_{12} & T_{22}
\end{array}\right)\ \ \
\ee
where we substituted the genus-one and genus-two relations:
\be
\xi^{(2)}_8(\tau) \ \stackrel{(\ref{gen1CDGG})}{=} \
-2\xi^{(0)}_8(\tau) + 3\xi^{(1)}_8(\tau),\nn\\
\xi^{(3)}_8(\tau)  \ \stackrel{(\ref{gen2CDGG})}{=} \
-6\xi^{(0)}_8(\tau) + 7\xi^{(1)}_8(\tau)
\label{gen12CDGG}
\ee
and
\be
\xi^{(3)}_8\left(\begin{array}{cc} T_{11} & T_{12}\\T_{12} & T_{22}
\end{array}\right) = 8\xi^{(0)}_8
\left(\begin{array}{cc} T_{11} & T_{12}\\T_{12} & T_{22}
\end{array}\right)
-14\xi^{(1)}_8
\left(\begin{array}{cc} T_{11} & T_{12}\\T_{12} & T_{22}
\end{array}\right)
+ 7\xi^{(2)}_8
\left(\begin{array}{cc} T_{11} & T_{12}\\T_{12} & T_{22}
\end{array}\right)
\ee
We omit characteristics labels in this section
to simplify the formulas.

Similarly, to check the decomposition with $g = g_1+g_2$,
\be
\Xi_8 = \sum_{p=0}^g h_p \xi^{(\!p)}_8  \longrightarrow
\underline{\sum_{p=0}^g h\!_p \xi^{(\!p)}_8\otimes \xi^{(p)}_8}
= \left(\sum_{p=0}^{g_1} h\!_p \xi^{(\!p)}_8\right)\otimes
\left(\sum_{p=0}^{g_2} h\!_p \xi^{(\!p)}_8\right) =
\Xi_8\otimes \Xi_8
\ee
one needs to know the analogues of (\ref{gen1CDGG}) and
(\ref{gen2CDGG}) to substitute into the underlined expression.
After that the next equality is just an algebraic identity
for coefficients $h\!_p$ in (\ref{hps}).
Remarkably, generalizations of (\ref{gen1CDGG})  and (\ref{gen2CDGG})
can be found for all genera by pure algebraic means:
analyzing restrictions to hyperelliptic loci. Despite these loci
have high codimension $g-2$, all the coefficients are
unambiguously fixed in these restrictions.
Eqs.(\ref{gen1CDGG})  and (\ref{gen2CDGG}) themselves are actually
enough to validate decompositions $g = m\cdot 1 + n\cdot 2$
with various $m$ and $n$.

To show just one more example, the decomposition
$4\rightarrow 2+2$ implies that
\be
H_0\xi^{(0)}\otimes \xi^{(0)} + H_1\xi^{(1)}\otimes \xi^{(1)}
+ H_2\xi^{(2)}\otimes \xi^{(2)}
+ H_3\xi^{(3)}\otimes \xi^{(3)}
+ H_4 \xi^{(4)}\otimes \xi^{(4)} = \nn \\
= \Big(h_0\xi^{(0)} + h_1\xi^{(1)} + h_2\xi^{(2)}\Big)\otimes
\Big(h_0\xi^{(0)} + h_1\xi^{(1)} + h_2\xi^{(2)}\Big)
\label{Hhrels1}
\ee
where $H_p$ correspond to genus $4$ (the forth line in
(\ref{hps}) while $h_p$ -- to genus $2$ (the second line in
(\ref{hps}),-- and genus-two modular forms $\xi^{(p)}_8[e]$
are related by (\ref{gen2CDGG}):
\be
\xi^{(3)} = 8\xi^{(0)} - 14 \xi^{(1)} + 7\xi^{(2)}, \nn \\
\xi^{(4)} = 56\xi^{(0)} - 90 \xi^{(1)} + 35\xi^{(2)}
\ee
Collecting the coefficients at different independent products of
forms in (\ref{Hhrels1}), we obtain:
\be
\begin{array}{c|ccccccc|ccccccc}
&&&&&&&&&&&&&&\\
\xi^{(0)}\otimes\xi^{(0)}
& H_0 &+& 8^2 H_3 &+& 56^2H_4 &=& h_0^2 &
\frac{64}{315} &-& \frac{8^2}{168} &+& \frac{56^2}{5040}
&=& \frac{4}{9}\\ &&&&&&&&&&&&&&\\
\xi^{(1)}\otimes\xi^{(1)}
& H_1 &+& 14^2 H_3 &+& 90^2H_4 &=& h_1^2 &
-\frac{4}{21} &-& \frac{14^2}{168} &+& \frac{90^2}{5040}
&=& \frac{1}{4}\\ &&&&&&&&&&&&&&\\
\xi^{(2)}\otimes\xi^{(2)}
& H_2 &+& 7^2 H_3 &+& 35^2H_4 &=& h_2^2 &
\frac{1}{18} &-& \frac{7^2}{168} &+& \frac{35^2}{5040}
&=& \frac{1}{144}\\ &&&&&&&&&&&&&&\\
\xi^{(0)}\otimes\xi^{(1)}
& &-& 8\cdot 14 H_3 &-& 56\cdot 90H_4 &=& h_0h_1 &
&& \frac{112}{168} &-& \frac{56\cdot 90}{5040}
&=& -\frac{1}{3}\\ &&&&&&&&&&&&&&\\
\xi^{(0)}\otimes\xi^{(2)}
& && 8\cdot 7 H_3 &+& 56\cdot 35H_4 &=& h_0h_2 &
&-& \frac{56}{168} &+& \frac{56\cdot 35}{5040}
&=& \frac{1}{18}\\ &&&&&&&&&&&&&&\\
\xi^{(1)}\otimes\xi^{(2)} & &-& 7\cdot 14 H_3 &-& 90\cdot 35 H_4 &=& h_1h_2&
&& \frac{98}{168} &-& \frac{90\cdot 35}{5040}
&=& -\frac{1}{24}\\ &&&&&&&&&&&&&&\\
\end{array}
\ee
Equalities in the last column obtained by substitution
of the coefficients from (\ref{hps}) are indeed true.

\section{Conclusion}

To conclude, we reviewed spectacular new development in
perturbative superstring theory, caused by the ground-breaking
papers \cite{DHP1}-\cite{DHP9} of Eric D'Hoker and Duong Phong
and their direct continuation in \cite{Zh}-\cite{Ma}.
The main reason why these formulas have not been discovered
in the first attack on NSR measures in 1980's seems
related to three prejudices.

First, starting from \cite{MP1}, the vanishing of cosmological
constant was attributed to Riemann identities, while
the simple relation (\ref{chiRid}) at genus one allowed
two kinds of generalizations: to (\ref{Rid}) and to (\ref{chif}).
It turned out that the second choice is more appropriate.

Second, NSR measure $d\mu_e$ was believed to be proportional
to $\theta_e^4$, so that expressions for to 1,2,3,4-point
functions would not contain $\theta_e$ in denominators.
Remarkably, this prejudice was still alive in \cite{DHP1}
and was finally broken only in \cite{CPG1}, though it was
actually based on the misleading overestimate of the role
of the Riemann identities (since they had a generalization
(\ref{Ridz}), the vanishing of 1,2,3-point functions would
automatically come together with that of the 0-function
-- if Riemann identities were the right thing to rely upon).

Third, naive integration over odd supermoduli was associated
with a correlator of the superghost $\beta,\gamma$-fields \cite{M2},
which produced a non-trivial theta-function in {\it denominator}
and summation over spin structures (theta-characteristics)
looked hopeless.
An artistic choice of odd moduli was then required in order
to eliminate this theta-function and perform the summation.
Exact treatment of odd moduli in \cite{DHP1}-\cite{DHP9}
confirmed that
the measure $d\mu_e$ is simple and has nothing non-trivial
in denominator (at least for genus two) and this opened the
way for a new stage of guess-work, based on the search of the
modular forms with given properties.

Today all these problems seem to be largely resolved,
{\bf the outcome
-- eqs.(\ref{xior}), (\ref{hps}) and (\ref{NSRGru}) --}
is nearly obvious (once you know it)
and it {\bf deserves to be widely known}.
Our main goal in this text was to give as simple presentation
of the subject as possible, avoiding unnecessary details
about supermoduli integration and modular-forms theory,
relying instead only on widespread
knowledge of elementary string theory.
To avoid overloading the text we did not include consideration
of non-renormalization theorems for 1,2,3-point functions \cite{Mar},
in particular, the resolution of the $\theta_e^4$ "paradox", and
the most interesting expressions for 4-point functions
(found and proved in above-cited references).
Already at the level of 4-point functions the NSR string with GSO
projection can be compared to Green-Schwarz superstring
\cite{GSst}-\cite{KM}, where equally impressive progress
is also achieved in recent years due to the works of
Nathan Berkovits \cite{Berk} -- and this is a separate issue of
great importance to be addressed elsewhere.

\section*{Acknowledgements}

I am grateful to my colleagues and friends,
who taught me a lot about various aspects of Riemann surfaces
and string measures:
A.Beilinson, A.Belavin, A.Gerasimov, E.D'Hoker, R.Iengo,
R.Kallosh, I.Krichever, A.Levin, D.Lebedev, O.Lechten\-feld,
Yu.Manin, G.Moore, P.Nelson, M.Olshanetsky, D.Phong, G.Shabat,
A.Schwarz, T.Shiota, A.Turin, A.Voro\-nov, Al.Zamolod\-chikov
and especially to V.Knizhnik, A.Perelomov and A.Rosly.

This work  is partly supported by Russian Federal Nuclear
Energy Agency and Russian Academy of Sciences,
by the joint grant 06-01-92059-CE,  by NWO project 047.011.2004.026,
by INTAS grant 05-1000008-7865, by ANR-05-BLAN-0029-01 project,
by RFBR grant 07-02-00645 and
by the Russian President's Grant of Support for the Scientific
Schools NSh-3035.2008.2


\begin{thebibliography}{12}

\bibitem{BPZ} A.Belavin, A.Polyakov and A.Zamolodchikov,
{\it Infinite Conformal Symmetry in Two-Dimensional Quantum
Field Theory}, Nucl.Phys. {\bf B241} (1984) 333
\bibitem{PO} A.Polyakov, {\it Quantum Geometry of Bosonic String},
Phys.Lett. {\bf 103B} (1981) 207-210;
{\it Quantum Geometry of Fermionic String},
Phys.Lett. {\bf 103B} (1981) 211-213
\bibitem{BK} A.Belavin and V.Knizhnik,
{\it Algebraic Geometry and the Geometry of Quantum Strings},
Phys.Lett. {\bf 168B} (1986) 201-206;  ZhETF {\bf 91} (1986) 247
\bibitem{VV}
E.Verlinde and H.Verlinde,
{\it Chiral Bosonization, Determinants and the String Partition
Function}, Nucl. Phys. {\bf B288} (1987) 357-396
\bibitem{DHP0}
E.D'Hoker and D.Phong, Nucl.Phys. {\bf B292} (1987) 109;
{\it The Geometry of String Perturbation Theory}, Rev.Mod.Phys.,
{\bf 60} (1988) 917-1065
\bibitem{Krev} V.Knizhnik,
{\it Multiloop Amplitudes in the Theory of Quantum Strings
and Complex Geometry}, Usp. Phys. Nauk, {\bf 159} (1989) 401-453
\bibitem{GMMOS} A.Gerasimov, A.Marshakov, A.Morozov,
M.Olshanetsky and S.Shatashvili, {\it Wess-Zumino-Witten model
as a theory of free fields},
Int.J.Mod.Phys. {\bf A5} (1990) 2495-2589, sec.5
\bibitem{MPrev} A.Morozov and A.Perelomov, {\it Strings and
Complex Geometry}, in Modern Problems of Mathematics,
VINITI, Moscow, 1990;
Encyclopedia o Mathematical Sciences, {\bf 54} (1993) 197-280,
Springer
\bibitem{Shopar}
P.Di Veccia, M.Frau, A.Lerda and S.Sciuto, {\it A Simple Expression
for Multiloop Amplitude in the Bosonic String}, Phys.Lett.
{\bf 199B} (1987) 49; \\
A.Losev, {\it The Chiral Fermion Determinant in the Schottky
Parametrization}, JETP Lett. {\bf 49} (1989) 424-426;
{\it Calculation of Chiral Determinants and Multiloop Amplitudes
by Cutting and Sewing Method}, Phys.Lett. {\bf 226B} (1989) 67-72
\bibitem{Selt} N.Hurt, {\it Geometric Quantization in Action},
D.Reidel Publ.Company, 1983, sec.18; \\
M.Baranov and A.Schwarz, {\it On the Multiloop Contribution
to the String Theory}, Int.J.Mod.Phys. {\bf A2} (1987) 1773; \\
M.Baranov, I.Frolov, Yu.Manin and A.Schwarz, {\it A Superanalog
of the Selberg Trace Formula and Multiloop Contributions for
Fermionic Strings}, Comm.Math.Phys. {\bf 111} (1987) 373-392
\bibitem{Mum} D.Mumford, Mathematica, {\bf 12} \# 6 (1968) 67;
L'Ens.Math. {\bf 23} (1977) 39
\bibitem{YuM} Yu.Manin, Pis'ma v ZhETF, {\bf 43} (1986) 161 ; \\
A.Beilinson and Yu.Manin, {\it The Mumford form and the Polyakov
measure in string theory}, Comm. Math. Phys. {\bf 107} (1986) 359-376
\bibitem{superst}
J.Schwarz, {\it Superstring Theory}, Phys.Reports, {\bf 89}
(1982) 223-322; \\
M.Green and J.Schwarz, {\it Anomaly Cancelations in Supersymmetric
D=10 Gauge Theory and Superstring Theory}, Phys.Lett. {\bf 149B}
(1984) 117-122; \\
S.Mandelstam, {\it Interacting String Picture of the Fermionic
String}, Progr.Theor.Phys.Suppl. {\bf 80} (1986) 163; \\
M.Green, J.Schwarz and E.Witten, {\it Superstring Theory},
Cambridge University Press, 1987; \\
A.Polyakov, {\it Gauge Fields and Strings}, 1987; \\
J.Polchinsky, {\it String Theory}, Cambridge University Press, 1998
\bibitem{heter} D.Gross, J.Harvey, E.Martinec and R.Rohm,
{\it Heterotic String}, Phys.Rev.Lett. {\bf 54} (1985) 502-505;
{\it Heterotic string theory. I. The Free Heterotic String},
Nucl.Phys. {\bf B256} (1985) 253-284;
{\it Heterotic string theory. I. The Interactive Heterotic String},
Nucl.Phys. {\bf B267} (1986) 75-124
\bibitem{BKMP} A.Belavin, V.Knizhnik, A.Morozov and A.Perelomov,
{\it Two- and three-loop amplitudes in the bosonic string theory},
Phys.Lett. {\bf 177B} (1986) 324; Pis'ma v ZhETF, {\bf 43} (1986)
319
\bibitem{M1} A.Morozov,
{\it Explicit formulae for one, two, three and four loop string
amplitudes}, Phys.Lett. {\bf 184B} (1987) 171-176;
{\it Analytical Anomaly and Heterotic String in the Formalism
of Continual Integration}, Phys.Lett. {\bf 184B} (1987) 177-183
\bibitem{GM} G.Moore, {\it Modular Forms and Two-Loop String
Physics}, Phys.Lett. {\bf 176B} (1986) 69
\bibitem{DHP1} E.D'Hoker and D.Phong, {\it Two Loop Superstrings. I.
Main Formulas}, Phys.Lett. {\bf B529} (2002) 241-255, hep-th/0110247
\bibitem{DHP2} E.D'Hoker and D.Phong, {\it Two Loop Superstrings. II.
The Chiral Measure on Moduli Space},
Nucl.Phys. {\bf B636} (2002) 3-60, hep-th/0110283
\bibitem{DHP3} E.D'Hoker and D.Phong, {\it Two Loop Superstrings. III.
Slice Independence and Absense of Ambiguities},
Nucl.Phys. {\bf B636} (2002) 61-79, hep-th/0111016
\bibitem{DHP4} E.D'Hoker and D.Phong, {\it Two Loop Superstrings. IV.
The Cosmological Constant and Modular Forms},
Nucl.Phys. {\bf B639} (2002) 129-181, hep-th/0111040
\bibitem{DHP5} E.D'Hoker and D.Phong, {\it Asyzygies, Modular Forms
and the Superstring Measure. I
},  Nucl.Phys. {\bf B710} (2005) 58-82, hep-th/0411159
\bibitem{DHP6} E.D'Hoker and D.Phong, {\it Asyzygies, Modular Forms
and the Superstring Measure. II
},  Nucl.Phys. {\bf B710} (2005) 83-116, hep-th/0411182
\bibitem{DHP7} E.D'Hoker and D.Phong, {\it Two Loop Superstrings. V.
Gauge Slice Dependence of the N-Point Function}, Nucl.Phys.
{\bf B715} (2005), 91-119, hep-th/0501196
\bibitem{DHP8} E.D'Hoker and D.Phong, {\it Two Loop Superstrings. VI.
Non-Renormalization Theorems and the 4-Point Functions},
Nucl.Phys.
{\bf B715} (2005), 3-90, hep-th/0501197
\bibitem{DHP9} E.D'Hoker and D.Phong, {\it Two Loop Superstrings. VII.
Cohomology of Chiral Amplitudes}, arXiv: 0711.4314
\bibitem{ADHP} K.Aoki, E.D'Hoker and D.Phong, {\it Two-Loop
Superstring on Orbifold Compactifications},
Nucl.Phys. {\bf B688} (2004) 3-69, hep-th/0312181
\bibitem{DHP2lrev} E.D'Hoker and D.Phong, {\it Lectures on
Two-Loop Superstrings}, hep-th/0211111
\bibitem{DGHP} E.D'Hoker, M.Gutperle and D.Phong,
{\it Two-Loop Superstrings and S-Duality}, Nucl.Phys.
{\bf B722} (2005) 81-118, hep-th/0503180
\bibitem{DHP10} E.D'Hoker and D.Phong,
{\it Complex Geometry and Supergeometry}, hep-th/0512197
\bibitem{XZh} Z.-G.Xiao and C.-J.Zhu, {\it Factorization of
the Two-Loop Four-Point Amplitude in Superstring Theory Revisited},
JHEP {\bf 0506} (2005) 002, hep-th/0412018
\bibitem{Zh} C.-J.Zhu, {\it A Formula for Multi-Loop 4-Particle
Amplitude in Superstring Theory}, hep-th/0503001
\bibitem{Mat} M.Matone and R.Volpato, {\it Higher genus
superstring amplitudes from the geometry of moduli space},
Nucl.Phys. {\bf B732} (2006) 321-340, hep-th/0506231
\bibitem{CP} S.Cacciatori, and F.Dalla Piazza,
{\it Two loop superstring ampliudes and $S_6$ representations},
arXiv: 0707.0646
\bibitem{CPG1} S.Cacciatori, F.Dalla Piazza and B.van Geemen,
{\it Modular Forms and Three Loop Superstring Amplitudes},
arXiv: 0801.2543
\bibitem{Gru} S.Grushevsky, {\it Superstring Amplitudes
in Higher Genus}, arXiv: 0803.3469 v1
\bibitem{CPG2} S.Cacciatori, F.Dalla Piazza and B.van Geemen,
{\it Genus Four Superstring Measures}, arXiv: 0804.0457 v1
\bibitem{Ma} R.Salvati Manni, {\it Remarks on Superstring
Amplitudes in Higher Genus}, arXiv: 0804.0512 v1
\bibitem{PG} F.Dalla Piazza and B.van Geemen,
{\it Siegel Modular Forms and Finite Symplectic Groups},
arXiv: 0804.3769
\bibitem{Mar}
E.Martinec, Phys.Lett. {\bf 171B} (1986) 189;
Nucl.Phys. {\bf B281} (1986) 157
\bibitem{Kni}
V.Knizhnik, {\it Covariant Fermionic Vertex in Superstrings},
Phys.Lett. {\bf 160B} (1985) 403-407;
Phys.Lett. {\bf 178B} (1986) 21; Pis'ma v ZhETF, {\bf 46}
(1987) 8
\bibitem{AG} L.Alvarez-Gaume, J.Bost, G.Moore, P.Nelson and
C.Vafa, {\it Bosonisation on Higher Genus Riemann Surfaces},
Comm.Math.Phys. {\bf 112} (1987) 503;
{\it Modular forms and the cosmological constant},
Phys.Lett. {\bf 178B} (1986) 41
\bibitem{MP1} A.Morozov and A.Perelomov,
{\it On vanishing of vacuum energy for superstrings},
Phys.Lett. {\bf 183B} (1987) 296-299
\bibitem{AS} J.Atick and A.Sen, Phys.Lett. {\bf 186B} (1987) 319
\bibitem{BoI}
M.Bonini and R.Iengo, Phys.Lett. {\bf 191B} (1987) 56
\bibitem{BoNe} J.Bost and P.Nelson, Phys.Rev.Lett. {\bf 57}
(1986) 795
\bibitem{VeVe}
E.Verlinde and H.Verlinde, {\it Multiloop calculations in covariant
superstring theory}, Phys.Lett. {\bf 192B} (1987) 95;
{\rm see also} \cite{VV}
\bibitem{MooNe}
G.Moore and P.Nelson, Nucl.Phys. {\bf B295} (1987) 312
\bibitem{Vor}
A.Voronov, Funk.Anal.i ego Prilozh., {\bf 21} (1987) 312;
{\it A formula for the Mumford measure in superstring theory},
Func.Anal.Appl. {\bf 22} (1988) 139-140
\bibitem{MP87}
A.Morozov and A.Perelomov,
{\it Statistical Sums in Superstring Theory. Genus 2},
Phys.Lett. {\bf 197B} (1987) 115-118;
Pis'ma v ZhETF, {\bf 46} (1987) 125; see also \cite{M2} and \cite{att}
\bibitem{AGG} L.Alvarez-Gaume, C.Gomez, G.Moore and C.Vafa,
{\it Strings in the Operator Formalism},
Nucl.Phys. {\bf B311} (1988) 333
\bibitem{LePa} O.Lechtenfeld and A.Parkes, {\it On the Vanishing
of the genus 2 Superstring Vacuum Amplitude}, Phys.Lett. {\bf 202B}
(1988) 75;  {\it On Covariant Multiloop Superstring Amplitudes},
Nucl.Phys. {\bf B332} (1990) 39-82
\bibitem{AG2}
L.Alvarez-Gaume, C.Gomez, G.Moore, P.Nelson, and C.Vafa,
{\it Fermionic Strings in the Operator Formalism},
Nucl.Phys. {\bf B311} (1988) 333
\bibitem{GaI} E.Gava and R.Iengo {\it Modular Invariance and the
Two Loop Vanishing of the Cosmological Constant}, Phys.Lett.
{\bf 207B} (1988) 283
\bibitem{GN} S.Giddings and P.Nelson, {\it The Geometry of
super Riemann Surfaces}, Comm.Math.Phys. {\bf 116} (1988) 607
\bibitem{KLT} I.Koh, D.Lust and S.Theisen, {\it Factorization
Properties of Genus Two Bosonic and Fermionic String Partition
Functions}, Phys.Lett. {\bf B208} 433
\bibitem{Yas} O.Yasuda,
{\it Factorization of a two loop Four Point Superstring Amplitude},
Phys.Rev.Lett. {\bf 60} (1988) 1688, erratum ibid {\bf 61} (1988)
1678; {\it Multiloop Modular Invariance of D=10
Type II Superstring Theory}, Nucl.Phys. {\bf B318} (1989) 397
\bibitem{IeZ} R.Iengo and C.J.Zhu, {\it Notes on Non-Renormalization
Theorem in Superstring Theories}, Phys.Lett.{\bf 212B} (1988) 309;
{Two Loop Computation of the Four-Particle Amplitude in the
Heterotic String}, Phys.Lett. {\bf 212B} (1988) 313
\bibitem{LeLe} O.Lechtenfeld and W.Lerche, {\it On
Non-Renormalization Theorems for Four-Dimensional Superstrings},
Phys.Lett. {\bf 227B} (1989) 373
\bibitem{DHP89} E.D'Hoker and D.Phong, {\it Conformal scalar
fields and chiral splitting on super Riemann surfaces},
Comm.Math.Phys. {|bf 125} (1989) 469-513
\bibitem{Lech} O.Lechtenfeld,
{\it On Finiteness of the Superstring}, Nucl.Phys. {\bf B322} (1989)
82; {\it Factorization and modular
invariance of multiloop superstring amplitudes in the unitary gauge},
Nucl.Phys. {\bf B338} (1990) 403-414
\bibitem{VRS} A.Rosly, A.Schwarz and A.Voronov, {\it Superconformal
Geometry and String Theory}, Comm.Math.Phys. {\bf 120} (1989) 437
\bibitem{Park} A.Parkes, {\it The Two-Loop Superstring Vacuum
Amplitude and Canonical Divisors}, Phys.Lett. {\bf 217B} (1989) 458
\bibitem{DRS} S.Dolgikh, A.Rosly and A.Schwarz,
{\it Supermoduli Spaces}, Comm.Math.Phys. {\bf 135} (1990) 91-100
\bibitem{ASch}
A.Schwarz, {\it Geometry of Fermionic String}, Proc.of the Int.
Congress of Mathematicians, Kyoto, Japan (1990) 1378-1386
\bibitem{PePe} R.Pettorino and F.Pezzella, {\it On the (B,C)-System
Contribution to Superstring Amplitudes}, Phys.Lett. {\bf B255}
(1991) 223
\bibitem{Ort} T.Ortin, {\it The Genus 2 Heterotic String
Cosmological Constant}, Nucl.Phys. {\bf B387} (1992) 280
\bibitem{M2} A.Morozov, {\it Two-loop statsum of superstring},
Nucl.Phys. {\bf B303} (1988) 343
\bibitem{ARS} J.Atick, J.Rabin and A.Sen, {\it An ambiguity in
fermionic string perturbation theory}, {\bf B299} (1988) 279-294
\bibitem{MM} G.Moore and A.Morozov, {\it Some Remarks on
Two-Loop Superstring Calculations}, Nucl.Phys. {\bf B306} (1988)
387-404
\bibitem{AGMS} J.Atick, G.Moore and A.Sen,
{\it Catoptric Tadpoles}, Nucl.Phys. {\bf B307} (1988)
221-273;  {\it Some Global Issues in String Perturbation Theory},
Nucl.Phys. {\bf B308} (1988) 1
\bibitem{LN} H.La and P.Nelson, {\it Unambiguous fermionic string
amplitudes}, Phys.Rev.Lett. {\bf 63} (1989) 24-27
\bibitem{GSO} F.Gliozzi, J.Sherk and D.Olive,
{\it Supersymmetry, Supergravity Theories and the Dual Spinor
Model}, Nucl.Phys. {\bf B122} (1976) 253-290
\bibitem{att}
A.Morozov and A.Perelomov,
{\it A note on many-loop calculations for superstring in the
NSR formalism}, Phys.Lett. {\bf 209B} (1988) 473-476; \\
A.Morozov, {\it Hyperelliptic Statsums in Superstring Theory},
Phys.Lett. {\bf 198B} (1988) 333;
{\it Point-wise vanishing of two-loop contributions to 1,2,3-point
functions in the NSR formalism}, Nucl.Phys. {\bf B318} (1989) 137-152;
\ {\it Straightforward proof of Lechtenfeld's formula
for the $\beta,\gamma$-correlators}, Phys.Lett. {\bf 234B}
(1990) 15-17; \
{\it On the two-loop contribution to the superstring four-point
function}, Phys.Lett. {\bf 209B} (1988) 473-476; \\
see also references in these papers
\bibitem{Mumb}
D.Mumford, {\it Tata Lectures on Theta}, Progr.in Math.
{\bf 28, 43}, Birkhauser, 1983, 1984
\bibitem{Fay} J.Fay, {\it Theta functions on Riemann surfaces},
Lect.Notes Math. {\bf 352}, Springer, 1973
\bibitem{FaKr} H.Farkas and I.Kra, {\it Riemann Surfaces},
Springer, 1980
\bibitem{Cle}
C.H.Clemens, {\it A Scrapbook of Complex Curve Theory},
Plenum Press, New York and London, 1980
\bibitem{Kob} N.Koblitz, {\it Introduction to Elliptic Curves and
Modular Forms}, Springer, 1984
\bibitem{Ig} J.-I.Igusa,
{\it On the graded ring of theta-constants}, Amer.J.Math.
{\bf 86} (1964) 219-246; ibid. {\bf 84} (1962) 175; ibid.
{\bf 89} (1967) 817;
{\it Theta Functions}, Springer-Verlag, 1972;
{\it Schottky's invariant and quadratic forms},
E.B.Christoffel Int.Symp., Aachen (1981) 352-362
\bibitem{No} S.Novikov, {\it Periodic problem for Korteveg-de-Vries
equation}, Func.Anal.i Prilozh. {\bf 8} \# 3 (1974) 54-66
\bibitem{Shi} T.Shiota,
{\it Characterization of Jacobian varieties in terms of
soliton equations}, Invent.Math. {\bf 83} (1986) 333-382
\bibitem{Kri} I.Krichever, {\it Integrable Linear Equations and
the Riemann-Schottky Problem}, math/0504192;
{\it Characterizing Jacobians via Trisecants of the Kummer
Variety}, math/0605625
\bibitem{GruKri} S.Grushevsky and I.Krichever,
{\it Integrable discrete Schroedinger equation and a characterization
of Prym varieties by a pair of quadrisecants}, arXiv: 0705.2829
\bibitem{KriSh} I.Krichever and T.Shiota,
{\it Abelian Solutions of the KP Equation} ,arXiv: 0804.0274
\bibitem{UFN3}
A.Morozov,
{\it Integrability and Matrix Models}, Phys.Usp. {\bf 37}
(1994) 1-55, hep-th/9303139; hep-th/9502091
\bibitem{BMM} H.Braden, A.Mironov and A.Morozov,
{\it QCD, Wick's Theorem for KdV $\tau$-functions and the
String Equation}, Phys.Lett. {\bf 514B} (2001) 293-298,
hep-th/0105169
\bibitem{Welt} G.Welters, {\it A Criterion for Jacobi Varieties},
Ann.of Math. {\bf 120} (1984) 497-504
\bibitem{LM} D.Lebedev and A.Morozov, {\it Statistical
sums of strings on hyperelliptic surfaces}, Nucl.Phys. {\bf B302}
(1988) 163
\bibitem{UMS}
D.Friedan and S.Shenker, {\it The Integrable Analytic Geometry
of Quantum String}, Phys.Lett. {\bf 175B} (1986) 287;
Nucl.Phys. {\bf B281} (1987) 509=545; \\
N.Ishibashi, Y.Matsuo and H.Ooguri, {\it Soliton Equations
and Free Fermions on Riemann Surfaces}, Mod. Phys. Lett.
{\bf A2} (1987) 119;\\
L.Alvarez-Gaume, C.Gomez and C.Reina, {\it Loop Groups,
Grassmannians and String Theory}, Phys.Lett. {\bf 190B} (1987)
55-62; \\
A.Morozov, {\it String Theory and the Structure of Universal
Module Space}, Phys.Lett. {\bf 196B} (1987) 325; \\
A.Schwarz, {\it Fermionic String and Universal Moduli Space},
Nucl.Phys. {\bf B317} (1989) 323
\bibitem{open}
O.Alvarez, {\it Theory of Strings with Boundaries},
Nucl.Phys. {\bf B216} (1983) 125; \\
S.Carlip, {\it Sewing Closed String Amplitudes}, Phys.Lett.
{\bf 209B} (1988) 464; \\
S.Blau, S.Carlip, M.Clements, S.Della Pietra and V.Della Pietra,
{\it The String Amplitude on Surfaces with Boundaries and
Crosscaps}, Nucl.Phys. {\bf B301} (1988) 285-303; \\
A.Morozov and A.Rosly, {\it Statistical Sums for open and/or
non-oriented strings}, Phys.Lett. {\bf 195B} (1987) 554;
{\it Strings and open Riemann surfaces}, Nucl.Phys. {\bf B326}
(1989) 205-221
\bibitem{DJS} D.-J.Smit, Comm.Math.Phys. {bf 111} (1987) 658
\bibitem{LevM} A.Levin and A.Morozov, {\it On the Foundations
of the Random Lattice Approach to Quantum Gravity}, Phys. Lett.
{\bf 243B} (1990) 207-214
\bibitem{GSst} M.Green and J.Schwarz,
{\it Covariant Description of Supesrstrings}, Phys.Lett.
{\bf 136B} (1984) 367-370
\bibitem{Car} S.Carlip, Nucl.Phys. {\bf B284} (1987) 365;
Phys.Lett. {\bf 186B} (1987) 141
\bibitem{KM} R.Kallosh and A.Morozov, {\it Green-Schwarz action
and loop calculations for superstring}, Int.J.Mod.Phys. {\bf A3}
(1988) 1943-1958;
{\it On the vanishing of multiloop contributions to 0,1,2,3-point
functions in the Green-Schwarz formalism for heterotic string},
Phys.Lett. {\bf 207B} (1988) 164-168
\bibitem{Berk} N.Berkovits, {\it Explaining the Pure Spinor
Formalism for the Superstring}, arXiv: 0712.0324 and references
therein

\end{thebibliography}
\end{document}